# Structure and relaxor ferroelectric behavior of novel tungsten bronze type ceramic $Sr_5BiTi_3Nb_7O_{30}$

Qiuwei He[1], Siegbert Schmid[2*], Xue Chen[3], Biaolin Peng[4], ChunChun Li[1], Changzheng Hu[1], Laijun Liu[1*], Manuel Hinterstein[5]

[1]*Guangxi Key Laboratory of Optical and Electronic Materials and Devices, Guilin University of Technology, Guilin 541004, China*

[2]*School of Chemistry, The University of Sydney, Sydney, NSW 2006, Australia.*

[3]*Guangxi Key Laboratory of Information Materials, Guilin University of Electronic Technology, Guilin 541004, China.*

[4]*School of Advanced Materials and Nanotechnology, Xidian University, Xi'an 710071, China*

[5]*Institute for Applied Materials - Ceramic Materials and Technologies (IAM - CMT), Karlsruhe Institute of Technology, 76131 Karlsruhe, Germany*

**Abstract:** A novel lead-free tungsten bronze type ceramic $Sr_5BiTi_3Nb_7O_{30}$, was prepared by a conventional solid-state reaction route. The room-temperature crystal structure shows an average structure with centro-symmetric space group *P*4/*mbm* identified by synchrotron XRD. Temperature dependence of dielectric permittivity indicates that $Sr_5BiTi_3Nb_7O_{30}$ is a ferroelectric relaxor with $T_m$ near 260 K. The ceramic displays stronger frequency dispersion and lower phase-transition temperature compared with $Sr_6Ti_2Nb_8O_{30}$. A macroscopic and phenomenological statistical model was employed to describe the temperature dependence of their dielectric responses. The calculated size of polar nanoregions (PNRs) of $Sr_5BiTi_3Nb_7O_{30}$ compared with $Sr_6Ti_2Nb_8O_{30}$ implies that the stronger diffusion phase transition for the former is related to the disorder emerged in both A and B sites. The smaller PNRs can be activated at lower temperature but have smaller electrical dipole moment. This is the origin of relaxor behavior of $Sr_5BiTi_3Nb_7O_{30}$ with lower $T_m$ and dielectric

---

[1] Corresponding authors.
E-mail addresses: siegbert.schmid@sydney.edu.au (S. Schmid), ljliu2@163.com (L. Liu)

permittivity. The PNRs is related to a local structure with a polar space group *P4bm*, which contributes to the dielectric frequency dispersion of relaxor behavior. This work opens up a promising feasible route to the development of relaxor ferroelectrics in tungsten bronze type oxides.

# 1. Introduction

Tungsten bronze (TB) type ferroelectrics, $A1_2A2_4C_4B1_2B2_8O_{30}$, with diffuse phase transition (DPT) behavior are known for their extraordinary physical properties, such as high electro-optic, piezoelectric and nonlinear optical efficiencies and have been investigated for years.[1,2] The structure consists of a network of corner-sharing $BO_6$ octahedra, which form tetragonal ($A1$) and pentagonal ($A2$) channels occupied by various distinct metal cations, respectively. Smaller triangular ($C$) channels are filled (or partially filled) by small low-charged cations such as, *e.g.*, $Li^+$ ions.[3]

$Sr_6Ti_2Nb_6O_{30}$ (STN) is one such tungsten bronze type ferroelectric, whose structure was investigated extensively. It was reported with tetragonal *P4bm* symmetry,[4] *Cmm*2 symmetry[5] and *Pba*2 space-group symmetry.[6] The non-centrosymmetric space group *P4bm* was first proposed by Ainger *et al.*[4], and confirmed in a number of publications.[7–9] However, later work by Whittle *et al.*[10] suggested that STN forms with $Pna2_1$ symmetry and a ~12.36×12.40×7.76 Å$^3$ unit cell instead. This was further supported by group theoretical considerations leading to the development of a general algorithm to determine possible distortions and space groups in tungsten bronze type compounds.[11–13]

Recently, the relaxor and ferroelectric behavior have been reported for $M_5RTi_3Nb_7O_{30}$ structures ($M$ = Ba and Sr; $R$ represents the larger rare earth ions ranging from La to Dy and also Bi).[14–16] In $M_5RTi_3Nb_7O_{30}$ tungsten bronzes, the relaxor behavior originates from disordered distribution of $M$ and $R$ ions on the $A1$ sites (tetragonal) and small size difference between $M$ and $R$ ions.

However, the nature of the relaxor behavior in the quaternary tungsten bronzes and its relation with compositions are not yet clear. In order to describe the dielectric response, the Curie-Weiss law was proposed to describe the temperature dependence of dielectric permittivity. Unfortunately, it is only suitable for the temperature above $T_m$, which means it cannot account for the mechanism correctly. Several models[17–19] were also proposed to quantitatively describe the temperature dependent dielectric permittivity of diffused phase transition (DPT). One of the popular models is the Gaussian function with a mean value $T_0$ and a standard deviation $\delta_G$ to describe the character of DPT.[20,21] The derived expression is:

$$\frac{\varepsilon_0}{\varepsilon - \varepsilon_\infty} = exp[\frac{(T-T_0)^2}{2\delta_G}] \qquad (1)$$

here $\varepsilon_\infty$ is the contribution from electronic and ionic polarization, and $\varepsilon_0$ is a temperature and frequency-dependent parameter. Later, a new and simple empirical equation for a phenomenological description of $\varepsilon(T)$ near $T_m$ was proposed by Santos and Eiras[22]:

$$\frac{\varepsilon_m}{\varepsilon} = 1 + (\frac{T-T_m}{\Delta})^\xi \qquad (2)$$

Where $\xi=1$ indicates a "normal" ferroelectric phase transition, while $\xi=2$ represents a so-called "complete" DPT. $\Delta$ is considered as an empirical diffuseness parameter that indicates the degree of the DPT. The merit of this equation is that it provides a good fitting of experimental data at temperatures around and above the dielectric dispersion region. However, it has limitations in describing the entire temperature range of $\varepsilon(T)$ and illustrating non-symmetric peaks around $T_m$ in relaxor.

In this work, the structure of $Sr_5BiTi_3Nb_7O_{30}$ (SBTN) was determined from synchrotron X-ray diffraction and an approach based on the macroscopic and phenomenological statistical model was employed to analyze the dielectric response of $Sr_5BiTi_3Nb_7O_{30}$ ferroelectric ceramics around and above the dielectric dispersion region. The results were compared with those of $Sr_6Ti_2Nb_6O_{30}$ (STN)[8], suggesting that stronger frequency dispersion results from the disorder existing in both A and B sites.

## 2. Experimental procedures

$Sr_5BiTi_3Nb_7O_{30}$ ceramic was prepared by solid-state reaction technique, using high-purity $SrCO_3$, $Bi_2O_3$, $TiO_2$, $Nb_2O_5$ powders as starting materials. After ball milling, the mixtures were sintered in high-purity alumina crucibles at 1200 °C in air for 4 h, whereafter, the disks annealed at 1000 °C for 2 h to obtain highly dense ceramics. The diffraction pattern of $Sr_5BiTi_3Nb_7O_{30}$ was measured using high-resolution synchrotron X-ray radiation (λ = 0.413 420 Å) at the beamline MSPD at ALBA (Barcelona, Spain). The crystal lattice parameters were refined by the Rietveld method using the FULLPROF program. The morphology of sintered sample was evaluated using secondary electrons images of Field Emission Scanning Electron Microscopy, FE-SEM (Hitachi S-4800). Dielectric properties in the range of 210 K to 690 K were measured with TZDM-200-800. *P-E* hysteresis loop was obtained from TFANALYZER-2000. A Thermo Scientific DXR Raman microscope (Waltham, MA) controlled by the software (Thermo Scientific Omnic) for room temperature Raman measurements was employed. The dielectric properties of the sample were measured using Impedance Analyzer (Agilent 4294A) in the frequency range from 40 Hz to 110 MHz at room temperature.

## 3. Results and discussion

3.1 Crystal structure and microstructure

High-resolution synchrotron X-ray powder diffraction was utilized to investigate the crystal structure of SBTN ceramic. Zhu et al reported the structure of $Sr_4(La_{1-x}Sm_x)_2Ti_4Nb_6O_{30}$[23] belongs to space group *P4bm*, where $La^{3+}$ and $Sm^{3+}$ ions randomly distribute at *A*1 site and $Ti^{4+}$ and $Nb^{5+}$ ions randomly locate at B site. Li *et al.* found that $Sr_5RTi_3Nb_7O_{30}$[9,16] (R = La, Sm) belongs to a centrosymmetric space group *P4/mbm*. Furthermore, the Ba-containing tungsten bronzes[24] also belong to centrosymmetric space group *P4/mbm,* just the same as $Sr_5RTi_3Nb_7O_{30}$.

In our work, X-ray powder diffraction data of SBTN can be indexed as a tetragonal tungsten bronze structure, and refined with centrosymmetric space group *P4/mbm,* through the Rietveld refinement as shown in Fig. 1. Since the diffraction data were collected at room temperature above $T_m$ (the temperature corresponding to the maximum of dielectric permittivity), it should be a paraelectric phase with a centro-symmetry. All peaks are indexed with the tungsten bronze phase with $a = b = 12.34195(6)$ Å and $c = 3.87675(2)$ Å, except some weak peaks consistent with a cubic cell with $a=b=c= 4.25$ Å. A complete list of atomic coordinates for the refinement of SBTN against synchrotron X-ray powder diffraction data is provided in Table I.

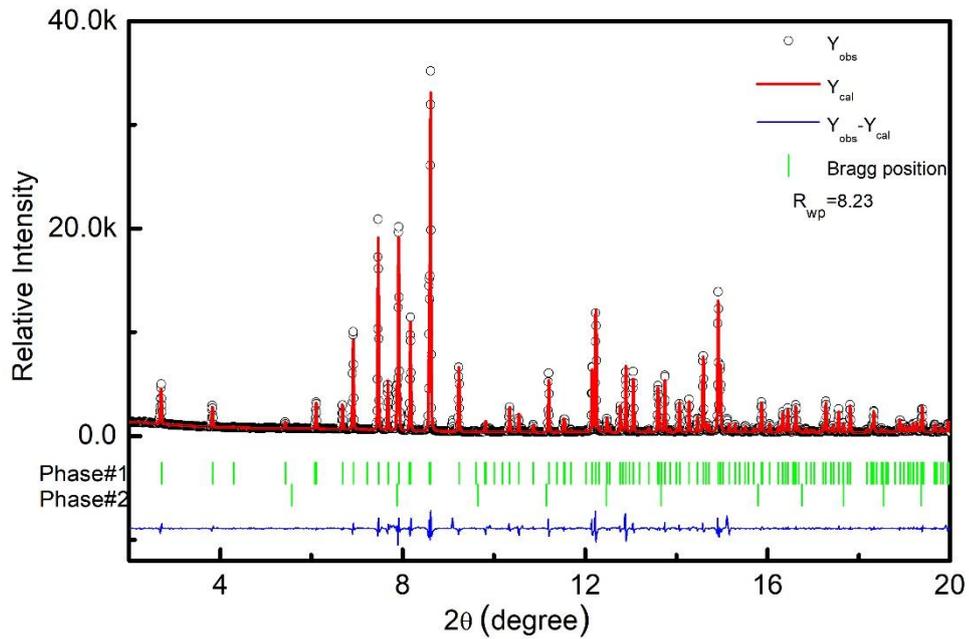

Fig. 1 The synchrotron X-ray powder diffraction date for SBTN (black) with a simulated pattern for the *P4/mbm* model (red) and the difference between them (blue). Allowed reflections are indicated with vertical green lines. Phase# 1 corresponds to the reflections of SBTN, phase# 2 is the reflections of an impurity phase indexed by a cubic symmetry with $a=b=c=4.25$ Å.

Table I. Atomic coordinates for SBTN resulting from refinement against synchrotron X-ray powder diffraction date

| Name | x | y | z | $U_{iso}$ |
|---|---|---|---|---|
| Bi1/Sr1 | 0 | 0 | 0 | 0.038 |
| Bi2/Sr2 | 0.1721 | 0.6721 | 0 | 0.038 |
| Ti1/Nb1 | 0 | 0.5 | 0.5 | 0.038 |
| Ti2/Nb2 | 0.0748 | 0.2159 | 0.5 | 0.038 |
| O1 | 0 | 0.5 | 0 | 0.038 |
| O2 | 0 | 0.279 | 0.779 | 0.038 |
| O3 | 0.064 | 0.218 | 0 | 0.038 |
| O4 | 0.345 | 0.007 | 0.5 | 0.038 |
| O5 | 0.146 | 0.066 | 0.5 | 0.038 |

Morphology of the sintered SBTN ceramic is shown in Fig. 2. The average grain size is about 3.8 μm and exhibits a compact characteristic. Most of grians are hexagon, suggesting the grains growth is complete.

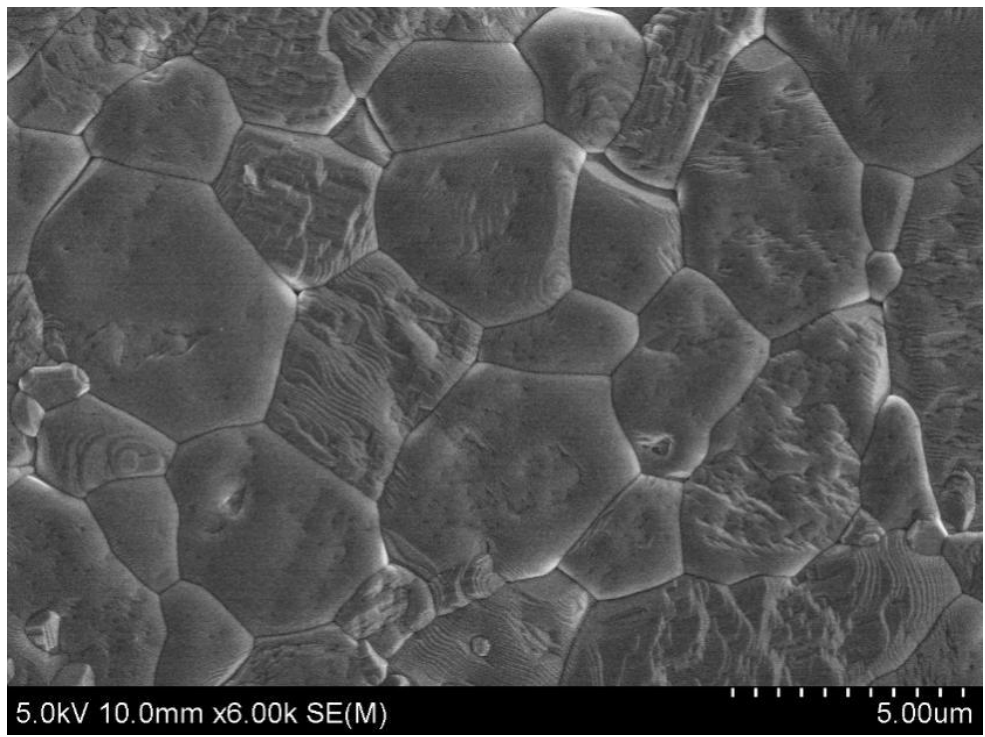

Fig. 2 SEM micrographs of the sintered $Sr_5BiTi_3Nb_7O_{30}$

3.2 Dielectric characterization

Temperature-dependence of the dielectric permittivity at frequencies of 500 Hz, 1 kHz, 10 kHz, 100 kHz and 1 MHz for SBTN is shown in Fig. 3. It is seen that the dielectric permittivity peaks at different frequencies exhibit a broad shape, which is similar to that of other tungsten-bronze compounds with relaxor behavior.[25] Moreover, the temperature corresponding to the maximum dielectric peak ($T_m$) moves towards higher temperature with increasing frequency. The SBTN ceramic shows a larger dielectric permittivity at 260 K, which may be attributed to thermally activated reorientation of dipole moments of polar nanoregions as well as the motions and interactions of the polar nanoregion boundaries.[26]

According to Zhu *et al.* independent of the cation distribution in the *A* site, larger difference in radius of A-site ions always causes normal ferroelectric transition.[27] In their study, $Sr_5RTi_3Nb_7O_{30}$ (R = La, Nd, Sm, Eu)[27] ceramics shows a larger difference in radius between the *A*1 and *A*2 sites, which reduces the overlap distribution between the two cation sites and enhances octahedral distortion, making ferroelectric transformation easier. Although a larger difference exists in *A* site in our work, the ceramic only exhibits relaxor-like behavior rather than a normal ferroelectric-paraelectric transition (Fig. 3), suggesting the dielectric anomaly may be dominated by the disorder in A and B sites. For tungsten bronze compounds, long-range dipolar coupling may be disrupted into polar clusters, which dominates the relaxation behavior at lower temperature. Considering the dielectric relaxation in SBTN, the disorder in A and B sites should be responsible for the removal of long-range dipolar coupling.

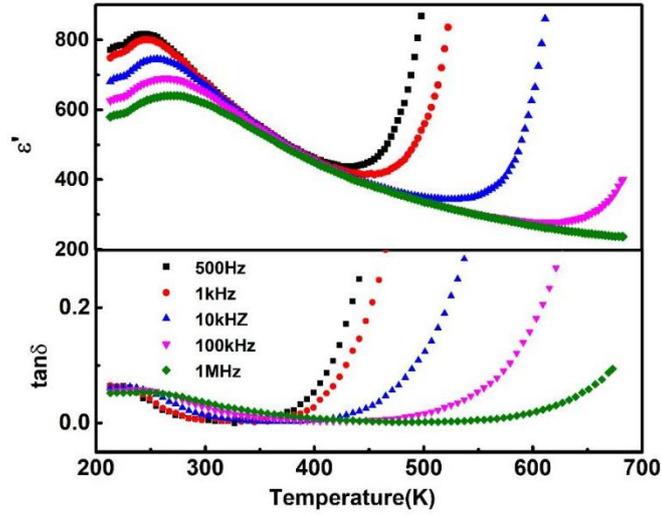

Fig. 3 Temperature dependence of dielectric permittivity and dielectric loss for SBTN

Room-temperature electric hysteresis loop of SBTN is shown in Fig. 4. The nonlinear hysteresis loop confirms the existence of spontaneous polarizations in the SBTN ceramic, which is consistent with a possible local structure with space group *P*4*bm* at room temperature. The slight loop agrees with the relaxor behavior, which is attributed to polar nanoregions (PNRs) rather than ferroelectric domains. A remnant polarization ($P_r$) of 0.95 μC/cm² and the corresponding coercive field ($E_c$) of 8.34 kV/cm are obtained at ambient temperature.

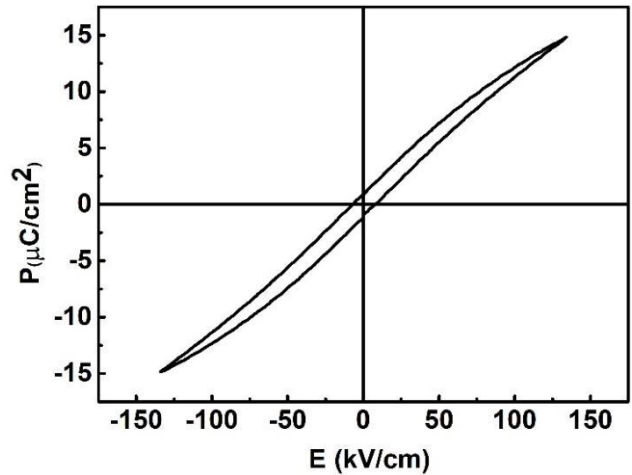

Fig. 4 Polarization (*P*) as function of applied electric field (*E*) measured at room temperature for SBTN

Raman spectroscopy of SBTN ceramic in the 60-1200 cm$^{-1}$ range is displayed in Fig. 5. According to the investigation of Raman spectra of tungsten bronze type compounds, the Raman bands around 260, 605, and 845 cm$^{-1}$ are considered as the three Raman characteristic peaks associated with the internal vibrational modes of the BO$_6$ octahedra, namely the O-B-O bending ($v_5$), O-B-O stretching ($v_2$), and B-O stretching ($v_1$) vibrational modes,[28] respectively. The Raman lines except those for the three characteristic vibrational modes can be associated with the external mode and are affected by cations in pentagonal and tetragonal tunnels. In Fig. 5 three characteristic vibrational modes are observed clearly in the Raman spectra of SBTN around 240 cm$^{-1}$ ($v_5$), 605 cm$^{-1}$ ($v_2$), and 845 cm$^{-1}$ ($v_1$), corresponding to internal vibrational modes of the (Ti/Nb)O$_6$ octahedra. The internal mode $v_2$ splits into two broad peaks at 515 cm$^{-1}$ and 605 cm$^{-1}$, and the mode $v_1$ also splits into two broad peaks at 833 cm$^{-1}$ and 913 cm$^{-1}$. Owning to the broad nature of the Raman lines, deconvolution is utilized to reduce each Raman vibration. The Gaussian profile is used as the Raman vibration fitting model as shown in Fig. 5. The room-temperature spectrum is deconvoluted into eight peaks, denoted as 1, 2, 3, …, 8. Peak 3 (240 cm$^{-1}$) belongs to the internal mode $v_5$, peak 5 (515 cm$^{-1}$) and 6 (605 cm$^{-1}$) belong to the internal mode $v_2$, whereas peak 7 (833 cm$^{-1}$) and 8 (913 cm$^{-1}$) belong to the internal mode $v_1$. The O-Ti/Nb-O stretching vibration $v_2$ shows an asymmetric shape, which deconvolutes into two vibrations at 515 cm$^{-1}$ (peak 5) and 605 cm$^{-1}$ (peak 6). Two aspects should be considered for the splitting of the internal modes: (1) the influence of external vibrations, which reflect the motions of the A-site cations; (2) the distorted octahedron and deviation from the ideal octahedral symmetry. In SBTN, the radii of Sr$^{2+}$ and Bi$^{3+}$ dominate the bonding requirements of the cations and determine the tilting type and distortion extent of the oxygen octahedra, which frustrates ferroelectric long-range order and result in the relaxor behavior.

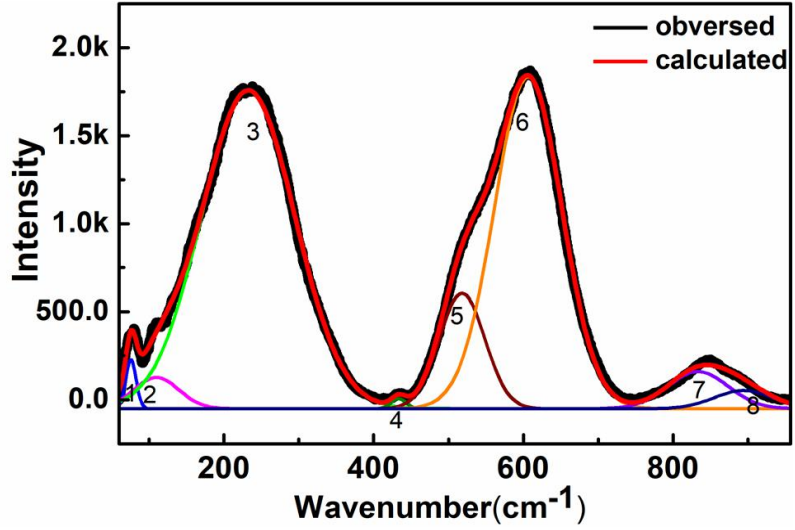

Fig. 5 Raman spectroscopy of SBTN over wave-number range of 60-1200 cm$^{-1}$. Red solid line: fitting of the deconvoluted Raman spectrum. The bottom lines include 8 peaks (marked as 1, 2, 3, …., 8)

3.3 Analysis of dielectric response

1) Dielectric dispersion model

Frequency dependence of the real part of permittivity $\varepsilon'$ and imaginary part of permittivity $\varepsilon''$ can be described by a combined the Universal Dielectric Response (UDR) and Cole-Cole equation[29]:

$$\varepsilon^* = \varepsilon' - i\varepsilon'' \qquad (3)$$

where

$$\varepsilon'_r = \varepsilon_\infty + \frac{(\varepsilon_s - \varepsilon_\infty)[1 + (\omega\tau)^{1-\alpha} \sin\left(\frac{\alpha\pi}{2}\right)]}{1 + 2(\omega\tau)^{1-\alpha} \sin\left(\frac{\alpha\pi}{2}\right) + (\omega\tau)^{2-2\alpha}} + \frac{\sigma_2}{\varepsilon_0 \omega^s} \qquad (4a)$$

where $\varepsilon_s$ is the static permittivity, $\varepsilon_\infty$ is the permittivity at very high frequency, $\omega$ is the angular frequency, $\tau$ is the mean relaxation time, $\alpha$ is the Cole-Cole parameter.

And

$$\varepsilon''_r = \frac{(\varepsilon_s - \varepsilon_\infty)(\omega\tau)^{1-\alpha} \cos\left(\frac{\alpha\pi}{2}\right)]}{1 + 2(\omega\tau)^{1-\alpha} \sin\left(\frac{\alpha\pi}{2}\right) + (\omega\tau)^{2-2\alpha}} + \frac{\sigma_1}{\varepsilon_0 \omega^s} \qquad (4b)$$

where s (0<s<1) is a constant, $\sigma_1$ (from free charge carrier) and $\sigma_2$ (from space charge) both are conductivity.

The room-temperature dielectric response can be described by Eqs. (4a) and (4b) as shown in Fig. 6. It is interesting to note that good agreement between experimental and calculated data over a wide frequency range for both $\varepsilon'$ and $\varepsilon''$ well supports the model. The obtained parameter $\alpha$ (0.713) is much smaller than 1, suggesting a strong interaction between PNRs, which is the nature of relaxor ferroelectrics rather than space charge polarization or charged carries hopping. The conductivity ($\sigma_1$) resulting from free charge carrier and the conductivity ($\sigma_2$) due to the space charges are 8.706 x$10^{-12}$ S and 9.977 x$10^{-11}$ S, respectively. Both of them are very low, indicating an intrinsic dielectric at room temperature.

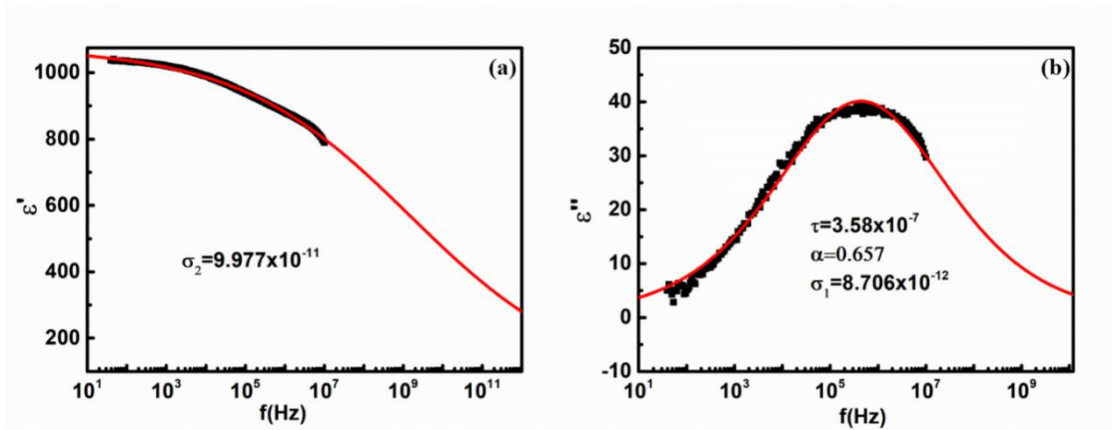

Fig. 6 Frequency dependence of $\varepsilon'$ (a) and $\varepsilon''$ (b) at room temperature. The red solid curves are the best fits to Eq. (4a) and Eq. (4b) for $\varepsilon'$ (a) and $\varepsilon''$ (b), respectively, which include both a Cole-Cole relaxation and an UDR contribution.

2) Fitting dielectric permittivity

To further probe the relaxor behavior, the degree of disorder or diffusivity ($\gamma$) can be obtained by the modified Curie–Weiss law[30,31]:

$$\ln\left(\frac{1}{\varepsilon} - \frac{1}{\varepsilon_m}\right) = \gamma \ln(T - T_m) + C' \qquad (5)$$

where $\varepsilon'_m$ is the dielectric permittivity at $T_m$, $C'$ is the Curie constant, $\gamma$ is an exponent indicating diffusivity, ranging from 1 (a normal ferroelectric) to 2 (an ideal relaxor ferroelectric). The fitting result of $\ln\left(\frac{1}{\varepsilon'} - \frac{1}{\varepsilon'_m}\right)$ as a function of $\ln(T - T_m)$ for SBTN at a frequency of 10 kHz is shown in Fig. 7(a), implying diffuse nature for SBTN. Fig. 7(b) confirms that the Curie–Weiss law $\frac{1}{\varepsilon} = \frac{T - T_{CW}}{C}$ ($T_{CW}$ is the Curie-Weiss temperature and C is a constant), only applies to temperatures much higher than $T_m$. The fitting parameters for SBTN are $C=1.41\times10^5$ and $T_{CW}=89.9$ K. The Curie constant with a magnitude of $10^5$ suggests SBTN to be a displacement-type ferroelectric.

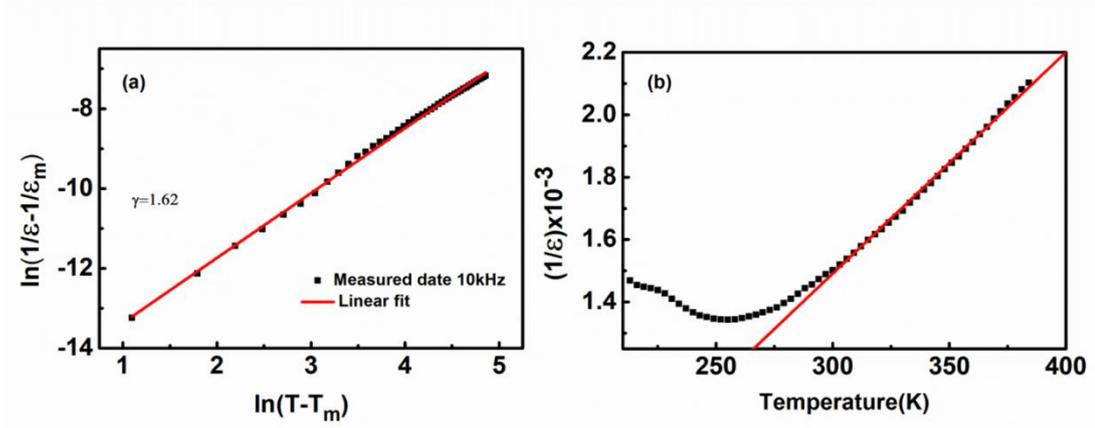

Fig. 7 Plot of $\ln(\frac{1}{\varepsilon'} - \frac{1}{\varepsilon'_m})$ as a function of $\ln(T - T_m)$ at 10 kHz (a) the temperature-dependence of the reciprocal dielectric permittivity of SBTN ceramic and (b) the fitting by the Curie-Weiss law at 10 kHz.

Recently, a macroscopic and phenomenological approach[32,33] was proposed to describe the dielectric response of ferroelectric relaxors in temperatures both below and above $T_m$. In such a model, individual dipoles which complies with the Maxwell-Boltzmann distribution at a given temperature are categorized into two groups. The first group consists of dipoles with lower kinetic energies ($[N_2(E_b,T)]$) than potential wells ($E_b$), whereas the second group ($[N_1(E_b,T)]$) can escape the wells and have more freedom in terms of orientations. It is not difficult to imagine that the two distinguishing dipoles contribute significantly different amounts to the dielectric response of ferroelectric relaxors. The total number of dipoles($N$) is equal to the sum of two groups. The relationship between those parameters is given by the following expressions:

$$N_1(E_b, T) = N\sqrt{\frac{4}{\pi}\sqrt{\frac{E_b}{k_B T}}} exp\left(-\frac{E_b}{k_B T}\right) + N erfc\sqrt{\frac{E_b}{k_B T}} \quad (6)$$

$$N_2(E_b, T) = N - N_1(E_b, T) \quad (7)$$

where $k_B$ is the Boltzmann constant, $T$ is the absolute temperature, and $erfc$ is the complementary error function. The total susceptibility is given by:

$$\chi(T, \omega) = \chi_1(T, \omega)P_1(E_b, T) + \chi_2(T, \omega)P_2(E_b, T) \quad (8)$$

here $\chi_1(T, \omega)$ and $\chi_2(T, \omega)$ are responses that correspond to $N_1(E_b, T)$ and $N_2(E_b, T)$, respectively. $\omega$ is the measurement frequency, and $P_1(E_b, T) = \frac{N_1(E_b, T)}{N}$, $P_2(E_b, T) = \frac{N_2(E_b, T)}{N}$.

According to this model, the temperature dependence of dielectric permittivity is proposed as follow[32]:

$$\varepsilon(T) = \frac{\varepsilon_1}{1 + b exp(-\frac{\theta}{T})} P_1(E_b, T) + \varepsilon_2 P_2(E_b, T) \quad (9)$$

where $\varepsilon_1$, $\varepsilon_2$, $b$ and $\theta$ are constants at a given frequency. We applied Eq. (9) to describe the dielectric response associated with polar nanoregions (PNRs) in SBTN ceramic and $Sr_6Ti_2Nb_8O_{30}$ (STN)[8] ceramic to better understand the ferroelectric relaxor nature. Clearly, both of them show satisfactorily fitting results in Fig. 8, which means that it is an effective way to describe the dielectric response of ferroelectric relaxors over the whole temperature range. The genesis of $E_b$ can be traced back to the random field attributed to cation disorder. The $E_b$ of SBTN and STN are 0.071 eV and 0.134 eV, respectively. According to Landau-Devonshire,[34,35] $\Delta G$ represents the energy barrier density for reorientation of a PNR ranging from $10^5 \sim 10^7$ J m$^{-3}$. $E_b$ can describe as $\Delta G * V$, $V$ represents the volume of a single PNR. The volume of a PNR for STN and SBTN is calculated to be 2.15 × $10^{-27}$ m³ and 1.04 × $10^{-27}$ m³, respectively. Corresponding the calculated size of PNRs ($\sqrt[3]{V}$) for STN and SBTN are 1.29 nm and 1.01 nm, respectively, which suggests that SBTN has smaller polar clusters due to the disorder in both A and B sites.

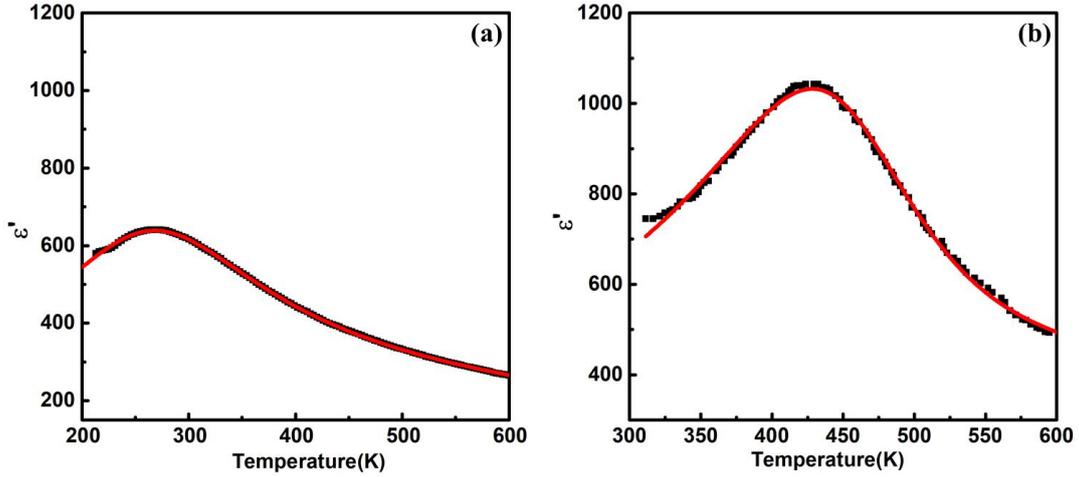

Fig. 8 Fit of dielectric permittivity obtained at 1M Hz for SBTN ceramic (a) and STN ceramic (b)

We show the $1/[1 + bexp\left(-\frac{\theta}{T}\right)]$ and the Maxwell-Boltzmann distribution versus temperature of SBTN and STN in Fig. 9. $P_1(E_b, T)$ and $P_2(E_b, T)$ have opposite trend as shown in Fig. 9(a) and Fig. 9(b). The number of dipoles that overcome the potential confinement ($P_1$) steadily increases with temperature, the number of dipoles with lower kinetic energies ($P_2$) decrease with temperature. Furthermore, SBTN has a greater number of dipoles that overcome potential wells than STN at room temperature. It suggests that the dielectric response of ferroelectric relaxor may be related to the active dipoles that overcome the potential wells. The dielectric permittivity is related not only to the number of active PNRs but also to the size of active PNRs (electrical dipole moment). Due to higher disorder in SBTN, the size of PNRs is smaller than that of STN, as a result, the dielectric permittivity is lower than that of later. The diffuseness exponent of the phase transition, $\gamma$, is calculated to be 1.42 and 1.62 for STN and SBTN, respectively, which confirms that the dielectric anomaly of SBTN could originate from the disorder at both A and B sites.

It has been stated that when $\omega_1$ is close to 1 in Fig. 9(c), the corresponding temperature is the freezing temperature $T_f$.[33] The value of STN is close to 1 around 300 K, but drops to zero at high temperature (~600 K), resembling the Fermi-Dirac function. Due to limited temperature measurement range of dielectric permittivity, the $T_f$ of SBTN is below 200K. The PNRs of SBTN can be activated at lower temperature due to smaller size. As the temperature increases to higher than $T_f$, the value of $\omega_1$ monotonously decreases. Although the structure of SBTN belongs to centrosymmetric space group *P*4/*mbm* at room temperature, a non-centrosymmetric, could be *P*4*bm*, may exit in a local structure (PNRs), which contributes to the dielectric frequency dispersion of relaxor behavior.

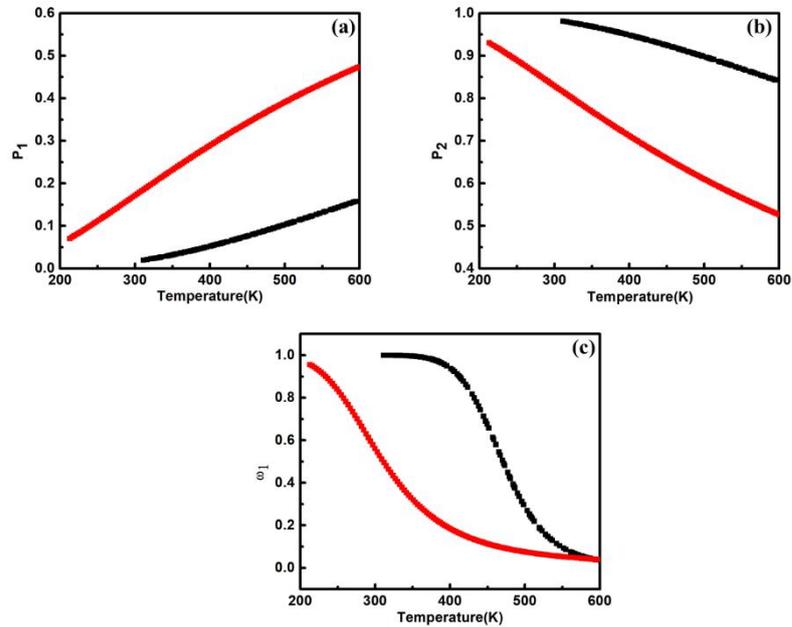

Fig. 9 The Maxwell-Boltzmann distribution versus temperature of SBTN (red line) and STN (black line) is shown in (a) and (b), and the $1/[1 + b\exp\left(-\frac{\theta}{T}\right)]$ versus temperature of SBTN (red line) and STN (black line) is shown in (c).

## 4. Conclusions

A new niobate tungsten bronze structure, $Sr_5BiTi_3Nb_7O_{30}$, was synthesized by solid state reaction and fully characterized for the first time. The crystal structure identified using synchrotron X-ray diffraction data shows an average structure with the centrosymmetric space group *P*4/*mbm* and a possible local structure with space group *P*4*bm* at room temperature. The temperature dependence of dielectric permittivity and dielectric loss of $Sr_5BiTi_3Nb_7O_{30}$ indicates relaxor ferroelectric nature with strong frequency dispersion with $T_m$ 260 K. The dielectric anomaly in $Sr_5BiTi_3Nb_7O_{30}$ is associated with the ionic disorder appeared at both A and B sites. The dielectric response associated with polar nanoregions (PNRs) at temperatures both below and above $T_m$ can be described by the statistical model. The size of PNRs is 1.01 nm, which is smaller than that of $Sr_6Ti_2Nb_8O_{30}$. The smaller PNRs can be activated at lower temperature but have smaller electrical dipole moment. It is the origin of relaxor behavior of $Sr_5BiTi_3Nb_7O_{30}$ with lower $T_m$ and dielectric permittivity.

**AUTHOR DECLARATIONS**

**Conflict of Interest**

The authors have no conflicts to disclose.

**DATA AVAILABILITY**

The data that support the findings of this study are available from the corresponding author upon reasonable request.

**ACKNOWLEDGMENTS**

The authors gratefully acknowledge the financial support from the Natural Science Foundation of

# References

[1] P. V. Bijumon, V. Kohli, O. Parkash, M. R. Varma, and M. T. Sebastian, Mater. Sci. Eng. B. **113**, 13(2004).

[2] Z. Han and P. Zhang, J. Chin. Ceram. Soc. **21,** 376(1993).

[3] A. Rotaru, D. C. Arnold, A. Daoud-Aladine, and F. D. Morrison, Phys. Rev. B. **83**, 184302(2011).

[4] F. W. Ainger, W. P. Brickley, and G. V. Smith, Br. Ceram. Proc. **18**, 221(1970).

[5] R. R. Neurgaonkar, J. G. Nelson, and J. R. Oliver, Mater. Res. Bull. **27**, 677(1992).

[6] E. O. Chi, A. Gandini, K. M. Ok, L. Zhang, and P. S. Halasyamani, Cheminform. **16,** 3616(2004).

[7] T. Ikeda and T. Haraguchi, Jpn. J. Appl. Phys. **9**, 422(1970).

[8] Y. Yuan, X. M. Chen, and Y. J. Wu, J. Appl. Phys. **98**, 2703(2005).

[9] X. L. Zhu and X. M. Chen, J. Am. Ceram. Soc. **95**, 3185(2012).

[10] T. A. Whittle, W. R. Brant, R. L. Withers, Y. Liu, C. J. Howard, and S. Schmid, J. Mater. Chem. C. **6**, 8890(2018).

[11] T. A. Whittle, S. Schmid, and C. J. Howard, Acta Crystallogr., Sect. B: Struct. Sci., Cryst. Eng. Mater. **71**, 342(2015).

[12] T. A. Whittle, S. Schmid, and C. J. Howard. Acta Crystallogr., Sect. B: Struct. Sci., Cryst. Eng. Mater. **74**, 742(2018).

[13] B. Campbell, C. J. Howard, T. B. Averett, T. A. Whittle, S. Schmid, S. Machlus, C. Yost, and H. T. Stokes, Acta Crystallogr., Sect. Found. Adv. **74**, 408(2018).

[14] Z. Hui, J. Wuhan Univ. Technol., Mater. Sci. Ed. **18**, 29(2003).

[15] K. Li, X. L. Zhu, X. Q. Liu, and X. M. Chen, Appl. Phys. Lett. **100**, 5048(2012).

[16] X. L. Zhu, K. Li, M. Asif Rafiq, X. Q. Liu, and X. M. Chen, J. Appl. Phys. **114**, 1804(2013).

[17]A. A. Boko and Z. G. Ye, Solid State Commun. **116**, 105(2000).

[18]A. A. Bokov, Y. H. Bing, W. Chen, Z. G. Ye, S. A. Bogatina, I. P. Raevski, S. I. Raevskaya, and E. V. Sahkar, Phys. Rev. B. **68**, 575(2003).

[19]C. Lei, A. A. Bokov, and Z. G. Ye, Ferroelectrics. **339**, 129(2006).

[20]V. V. Kirillov and V. A. Isupov, Ferroelectrics. **5**, 3(1973).

[21]G. Smolenskii, J. Phys. Soc. Jpn. **28**, 26(1970).

[22]I. A. Santos and J. A. Eiras, J. Phys. Condens. Matter. **13**, 11733(2001).

[23]X. L. Zhu, Y. Bai, X. Q. Liu, and X. M. Chen, J. Appl. Phys. **110**, 667(2011).

[24]W. B. Feng, X. L. Zhu, X. Q. Liu, M. S. Fu, X. Ma, S. J. Jia, and X. M. Chen, J. Am. Ceram. Soc. **101**, 1623(2018).

[25]L. Fang, H. Zhang, and J. B. Yan, Chin. J. Inorg. Chem. **18**, 1131(2002).

[26]Z. Liu, Y. Yuan, P. Jaimeewong, H. Wu, W. Ren, and Z. G. Ye, Ferroelectrics. **534**, 42(2018).

[27]X. L. Zhu, X. Q. Liu, and X. M. Chen, J. Am. Ceram. Soc. **94**, 1829(2011).

[28]X. L. Zhu, X. M. Chen, X. Q. Liu, and X. G. Li, J. Appl. Phys. **105**, 1664(2009).

[29]Z. Abdelkafi, N. Abdelmoula, H. Khemakhem, O. Bidault, and M. Maglione, J. Appl. Phys. **100**, 323(2006).

[30]K. Uchino and S. Nomura, Ferroelectrics. **44**, 55(1982).

[31]S. M. Pilgrim, A. E. Sutherland, and S. R. Winzer, J. Am. Ceram. Soc. **73**, 3122(2010).

[32]J. Liu, F. Li, Y. Zeng, Z. Jiang, L. Liu, D. Wang, Z. G. Ye, and C. L. Jia, Phys. Rev. B. **96**, 054115(2017).

[33]L. J. Liu, S. K. Ren, J. Zhang, B. L. Peng, L. Fang, and D. W. Wang, J. Am. Ceram. Soc. **101**, 2408(2018).

[34]M. J. Haun, E. Furman, S. J. Jang, and L. E. Cross, Ferroelectrics. **99**, 63(1989).

[35]K. Y. Chen, W. Fu, J. Liu., T. X. Yan, Z. C. Lan, L. Fang, B. L. Peng, D. W. Wang, and L. J. Liu, J. Am. Ceram. Soc. **103**, 2859(2020).